\documentclass[pra,twocolumn,showpacs]{revtex4-1}
\usepackage{graphicx}
\usepackage{amsmath}
\pdfoutput=1

\begin{document}

\title{Atom optical shop testing of electrostatic lenses using an atom interferometer}
\author{Ivan Hromada$^{1}$, Raisa Trubko$^{2}$, William F. Holmgren$^{1}$, Maxwell D. Gregoire$^{1}$, and Alexander D.  Cronin$^{1,2}$}
\affiliation{$^{1}$Department of Physics, University of Arizona, Tucson, Arizona 85721, USA}
\affiliation{$^{2}$College of Optical Sciences, University of Arizona, Tucson, Arizona 85721, USA}

\date{\today}

\begin{abstract}
We used an atom interferometer for atom optical shop testing of lenses for atomic de Broglie waves. We measured focal lengths and spherical aberrations of electrostatic lenses in three independent ways based on contrast data, phase data, or calculations of de Broglie wavefront curvature. We report focal lengths of -2.5 km and -21.7 km with 5\% uncertainty for different lenses. All three methods give consistent results. Understanding how lenses magnify and distort atom interference fringes helps improve atom beam velocity measurements made with phase choppers [New J. Phys. 13, 115007 (2011)], which in turn will improve the accuracy of atomic polarizability measurements.
\end{abstract}

\pacs{03.75.Be, 37.25.+k, 03.75.-b, 03.75.Dg}

\maketitle

\section{Introduction}\label{sec1}

\begin{figure}[b]
    \includegraphics[width=8cm]{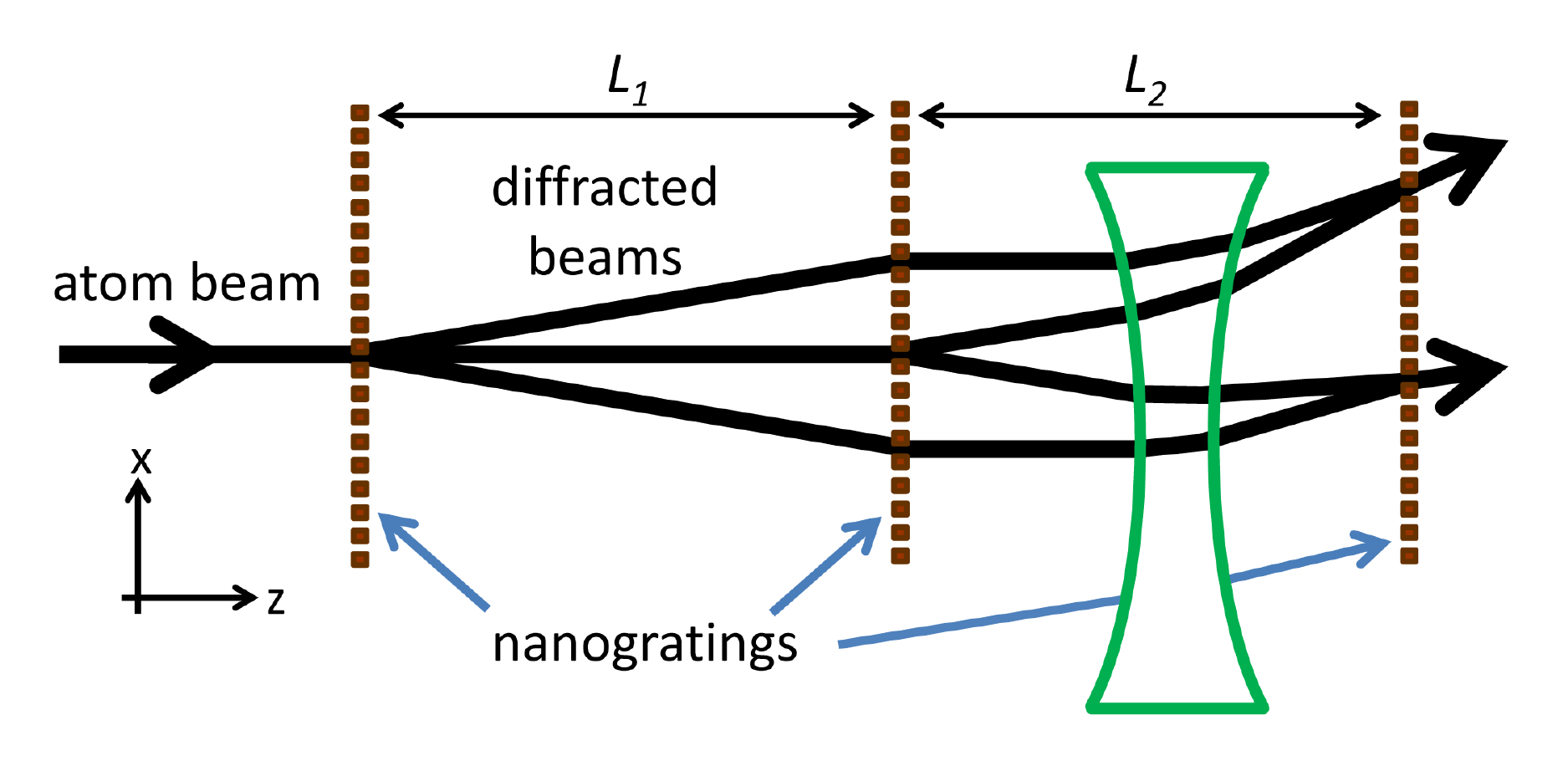}
    \caption{Schematic of a lens for atoms in an atom interferometer.  Atoms propagate in the $z$-direction and diffract from nano-gratings separated by distances $L_{1}$ and $L_{2}$. To study how the lens shifts, magnifies, and distorts atomic de Broglie wave interference fringes we vary $L_2$ (Sec.~\ref{sec4}) and translate the lens in the $x$-direction (Sec.~\ref{sec5}). $\hat{y}$ points out of the page. \label{fig:AIFM}}
\end{figure}

Atomic de Broglie waves can be focused, reflected, guided, and diffracted similarly to light waves \cite{BL89,ASM94,Meystre2001atom, folman2002microscopic, S09,CSJD09}, but the optical elements for atoms are necessarily different.  Lenses for atoms have been made from magnetic fields \cite{kaenders1996refractive,jardine2001hexapole, CLB06,ketterle1992trapping}, electric fields \cite{G55,noh2000imaging,KAG05}, zone plates \cite{CSSTM91,rdg00} and standing waves of radiation \cite{BFAP78,SPBM92,mcclelland1995atom}. Such lenses have been used for atom microscopes \cite{DGR99, KRS08} and for controlling the deposition of atoms on surfaces \cite{bam99,TBTCPB92,mcclelland1993laser}. In this paper, we show how a lens inside an atom interferometer can shift, magnify, and distort atom interference fringes.  In particular, we studied electrostatic lenses for neutral but polarizable atoms in a three-grating Mach-Zehnder atom interferometer as sketched in Fig.~\ref{fig:AIFM}.  Since interferometry is the gold standard for ordinary optical shop testing \cite{Malacara07,greivenkamp2004field,sasian2012introduction,goodwin2006field,hech98,pedrotti2008introduction}, we present this work as an example of atom optical shop testing.

This paper is also motivated by the aim to improve atom beam velocity measurements made with pulsing electrodes  known as \emph{phase choppers} \cite{HHKC11}. Ideally, a phase chopper would induce exactly $\pi$ radians phase shift for the fringes formed by any part of the atom beam.   The optics analogy for an ideal phase chopper is a non-dispersive prism, i.e. an optical component that causes beam deflection (fringe phase shift) independent of impact parameter or wavelength.  However, the electrodes that we developed for phase shifters do cause both a velocity-dependent and a position-dependent phase. Hence, a more realistic analogy for our phase choppers is a thin lens. To deflect atoms we use the lens off to one side of our atom beam with the lens optical axis parallel but not collinear to the atom beam. But even in this situation there is still dispersion and defocusing. We explored the consequences of velocity-dependence in \cite{HHKC11}, but the consequences of focusing were ignored until now.

We discovered that de Broglie wavefront curvature induced by phase choppers can cause systematic errors as large as 0.2\% in our atom beam velocity measurements. Furthermore, this can lead to 0.4\% errors in measurements of atomic polarizability. (For more examples of how atomic polarizability measurements made with atom interferometers also rely on measurements of atom beam velocity, see \cite{HRLC10,Eks95,Mif06,Ber07Arndt}.)  Correcting these errors is challenging because biases towards fast or slow atoms depend on several factors including the thickness of the atom beam, the location of the electrodes, misalignment of the gratings, and the size of the detector.  To understand how these factors are interrelated, we found it useful to model phase choppers as lenses inside an atom interferometer. Thus, we developed the optics analogy that electrostatic magnification can compensate for geometric magnification.

The rest of this paper is organized as follows: Sec. \ref{sec2} describes the electrostatic lens construction. In Sec.~\ref{sec3} we use simulations to discuss geometric magnification. In Sec.~\ref{sec4} and \ref{sec5} we use contrast and phase data to measure focal lengths and spherical aberration of lenses for atoms. In Sec.~\ref{sec6} we calculate focal lengths and aberrations theoretically.  Then, in Sec.~\ref{sec7}, we show how to include geometric magnification and electrostatic focusing when analyzing phase chopper data for velocity measurements. In Sec.~\ref{sec8} we prove that focusing effects are unavoidable when using electrodes with $y$-translation invariance.  Sec. \ref{sec9} is the discussion.

\section{Electrostatic lens construction}\label{sec2}

We built electrostatic lenses for atoms using two different electrode geometries, referred to here as Lens A and Lens B. Lens A is made from a charged cylinder near a grounded plane. Lens B is made from two parallel cylinders with equal diameters held at opposite voltages where the plane of symmetry between the cylinders remains at zero potential. Each assembly has a gap between the electrodes where atoms pass through inhomogeneous electric fields. Since the lenses work by virtue of polarizable atoms interacting with electric field gradients, the lenses can be turned on by applying a voltage (see Table~\ref{table1}) and turned off by grounding the electrodes. Table~\ref{table1} summarizes the electrode dimensions and resulting focal lengths in our experiment.

To cause uniform deflection over the height of our atom beam we orient the cylinders perpendicular to the atom beam velocity, and normal to the plane of our atom interferometer (parallel to $\hat{y}$ as defined in Fig.~\ref{fig:AIFM}). In this orientation the electrodes can deflect atoms by 50 nm, which results in a $\pi$ phase shift because we use 100-nm period gratings for our atom interferometer.  For comparison, several experiments \cite{HZ74,SMB74,SS08} have deflected atoms by much larger distances  (over 100 $\mu$m) with electrodes parallel to atom velocity.  The Arndt group \cite{berninger2007polarizability, Ste08} used a gap between the curved ends of two custom shaped electrodes to deflect molecules by several microns.  However, unlike our lenses, the electrodes used in \cite{HZ74,SMB74,SS08,berninger2007polarizability, Ste08} cause non-uniform deflection over the height of the beam.

Diagrams of electrodes in this geometry and calculations of the associated atom wave phase shifts have been presented before \cite{HHKC11,HRLC10,Rob04,Holm13,Kla11}. What is new here is the idea of using an atom interferometer to characterize these electrodes as lenses with focal lengths and aberrations.

\begin{table}[b]
\centering
\caption{Electrostatic lens dimensions and operating voltages.  Focal lengths ($f$) are calculated for 2000 m/s K atoms.}\label{table1}
\begin{ruledtabular}
        \begin{tabular}{l r r}
        Parameter & Lens A$_{2}$ & Lens B \\ [0.5ex]
        \hline
        cylinder radius ($R$)                 & 0.765 mm       & 6.350 mm         \\
        cylinder edge to symmetry plane ($a$)     & 0.893 mm        & 1.960 mm         \\
        electrode voltage ($V$)                  & - 3.0 kV          & $\pm$ 6.0 kV      \\
        paraxial focal length ($f$)            & -6.1 km        & -21.7 km          \\
        \end{tabular}
\end{ruledtabular}
\end{table}

Two copies of Lens A are installed in our atom interferometer as phase choppers. The first example of Lens A is located midway between the first two gratings.  We refer to it as Lens A$_{1}$.  The second one is located between the second and third gratings, where the cartoon of a lens in Fig.~1 is drawn.  We refer to this one as Lens A$_{2}$.  Lens B was built with a larger gap and larger cylinders as an interaction region for measurements of atomic polarizability. Lens B is located just in front of the second grating.  We need both the velocity measurements from the phase choppers (Lens A$_1$ and A$_2$) as well as phase shift measurements from Lens B in order to measure atomic polarizabilities.

\section{Fringe simulations}\label{sec3}

\begin{figure}[h]
\begin{center}
\includegraphics[width=9cm]{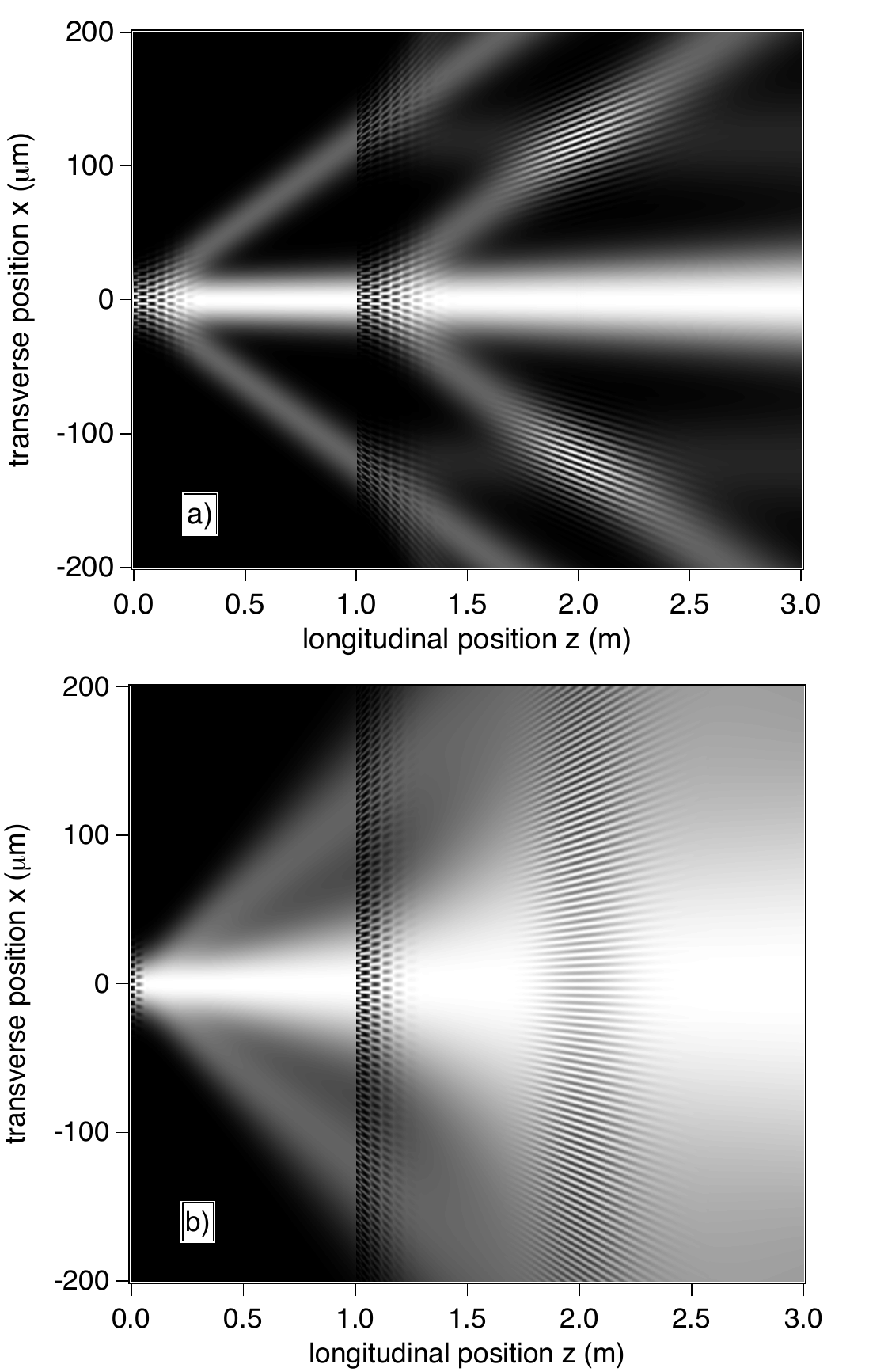}
\caption{Interference fringes formed after two gratings using a GSM beam simulation \cite{McM08}. The probability density $|\Psi|^{2}$ is plotted on a grey scale in the 0.4 mm x 3 m region. Transmission gratings are located at $z=0$ m and $z=1$ m.  The only difference between (a) and (b) is the transverse coherence of the GSM beam. In (a) fringes are formed in two distinct zones. In (b) the coherence length is shorter than the grating period, so the fringes appear to fan out in a range of directions, justifying the idea of geometric magnification ($M_\textrm{geom}$) defined in the text. $\hat{y}$ points out of the page. \label{fig:GSM}}
\end{center}
\end{figure}

To visualize how fringes are formed in an extended region, we present simulations made with Gaussian-Schell Model (GSM) beams.  In brief, a GSM beam is a mathematical ensemble of Gaussian beams with parameters for beam width and transverse coherence length \cite{friberg1982propagation,friberg1983spatial,gori1987bessel,McM08,McM08a}. Figure \ref{fig:GSM} shows the probability density for GSM beams as they propagate through two diffraction gratings located at $z$ = 0 and $z$ = 1 m. Table~\ref{table2} lists the parameters used in each simulation. The key idea is that only mutually coherent portions of the diffracted GSM beam components interfere.  Correlations between position and momentum of the beam components can then make interference fringes shift and spread out as a function of longitudinal position $z$.

Local structures in the fringe period, phase, and contrast relate to our experimental signals in subtle ways, which is why the simulations are helpful.  In the experiment, we average over much of this structure by using the third nanograting to moir\'{e} filter the fringes.  The signal thus comes from the ensemble of transmitted atoms that strike our 100-micron wide detector.

\begin{table}[b]
    \centering
\caption{Gaussian-Schell model (GSM) beam parameters (at $z=0$ m) used for the simulations shown in Fig.~2.  These parameters were selected to illustrate how fringes patterns are affected by the transverse coherence length of the incident beam.  For comparison, parameters that represent our experiment are also tabulated.} \label{table2}
    \begin{ruledtabular}
        \begin{tabular}{l r r r}
        Parameter &  Fig. 3(a) & Fig. 3(b) & Experiment\\ [0.5ex] 
        \hline
        de Broglie wavelength & 500 pm & 500 pm & 5 pm\\
        grating period & 5 $\mu$m & 5 $\mu$m & 100 nm\\
        coherence length & 25 $\mu$m & 2.5 $\mu$m & 50 nm \\
        beam width & 30 $\mu$m & 30 $\mu$m & 30 $\mu$m\\
        velocity ratio ($v_0/\sigma_v$) & 22 & 22 & 22 \\
        \end{tabular}
    \end{ruledtabular}
\end{table}

Figure \ref{fig:GSM}(a) shows resolved diffraction.   This occurs when a collimated beam has a transverse coherence length larger than the grating period.  Interference fringes are then found in two distinct regions, as suggested by the rays that depict two symmetric Mach-Zehnder interferometers in Fig.~\ref{fig:AIFM}.  Fringes in these separate regions are in phase at $z=2$ m but they shift away from each other (and become out of phase) as $z$ increases.  Thus, as the third grating is translated in the $z$-direction, moir\'{e} filtering can lead to reductions and revivals in contrast as a function of $\Delta L$,  the difference between grating separations (see Fig.~\ref{fig:AIFM}):
\begin{equation} \Delta L = L_2 - L_1. \end{equation}
Figure \ref{fig:GSM}(b) shows a simulation more representative of our experiments, in which the beam's initial transverse coherence length is slightly smaller than the grating period.  Hence, diffraction is poorly resolved and the fringes diverge in a several directions.  The way the fringe period changes with $z$ can be described by geometric magnification:
\begin{equation}
M_\textrm{geom} = \frac{\Delta L + 2L_1}{2L_1}.
\label{eqn:magnification}
\end{equation}
The fringe period is $d_{f} = M_\textrm{geom} d_{g}$, where $d_{g}$ is the grating period.  Geometric magnification occurs without a lens. It is a concept also found in studies of point-projection microscopy, the Lau effect \cite{patorski1989self}, and Talbot-Lau interferometry \cite{clauser1992new,BAZ03}.

For a more nuanced discussion, the local pitch and orientation of the fringes depends on the momenta of the interfering wave components.  Any two running waves with precise wave-vectors $\mathbf{k_1}$ and $\mathbf{k_2}$ make standing wave interference fringes with a wave-vector $\mathbf{k_f} = \mathbf{k_1} - \mathbf{k_2}$. But even monochromatic Gaussian and GSM beams contain \emph{distributions} of transverse momenta, and thus make fringes with a range of pitch so that different $\mathbf{k_f}$ can be observed in different locations. For collimated beams, fringes follow lines parallel to $ \mathbf{\overline{p}} = \hbar (\mathbf{k_1} + \mathbf{k_2})/2 $.  This is related to the the so-called \emph{separation phase} that was named by Dimopoulos \emph{et.}~\emph{al}.~in \cite{DGHK08} to describe how fringes shift as wave function components propagate through one another in a laboratory frame of reference.  GSM beams from a small aperture make fringes that follow hyperbolae in the x-z plane. GSM beams from wider apertures make interference visible over a more limited range in $z$, where fringe patterns from different parts of the source still overlap in phase, as in Talbot-Lau interferometers \cite{hornberger2012colloquium}. Thus, Eq.~(\ref{eqn:magnification}) is approximate, but works well enough to describe how the fringes diverge in our experiments. GSM beam simulations such as Fig.~\ref{fig:GSM}(b) justify this claim.  We also average over a velocity distribution in the simulations and the experiment.

\section{Using contrast to measure $f$} \label{sec4}

\begin{figure}[h]
\begin{center}
\includegraphics[scale=0.4]{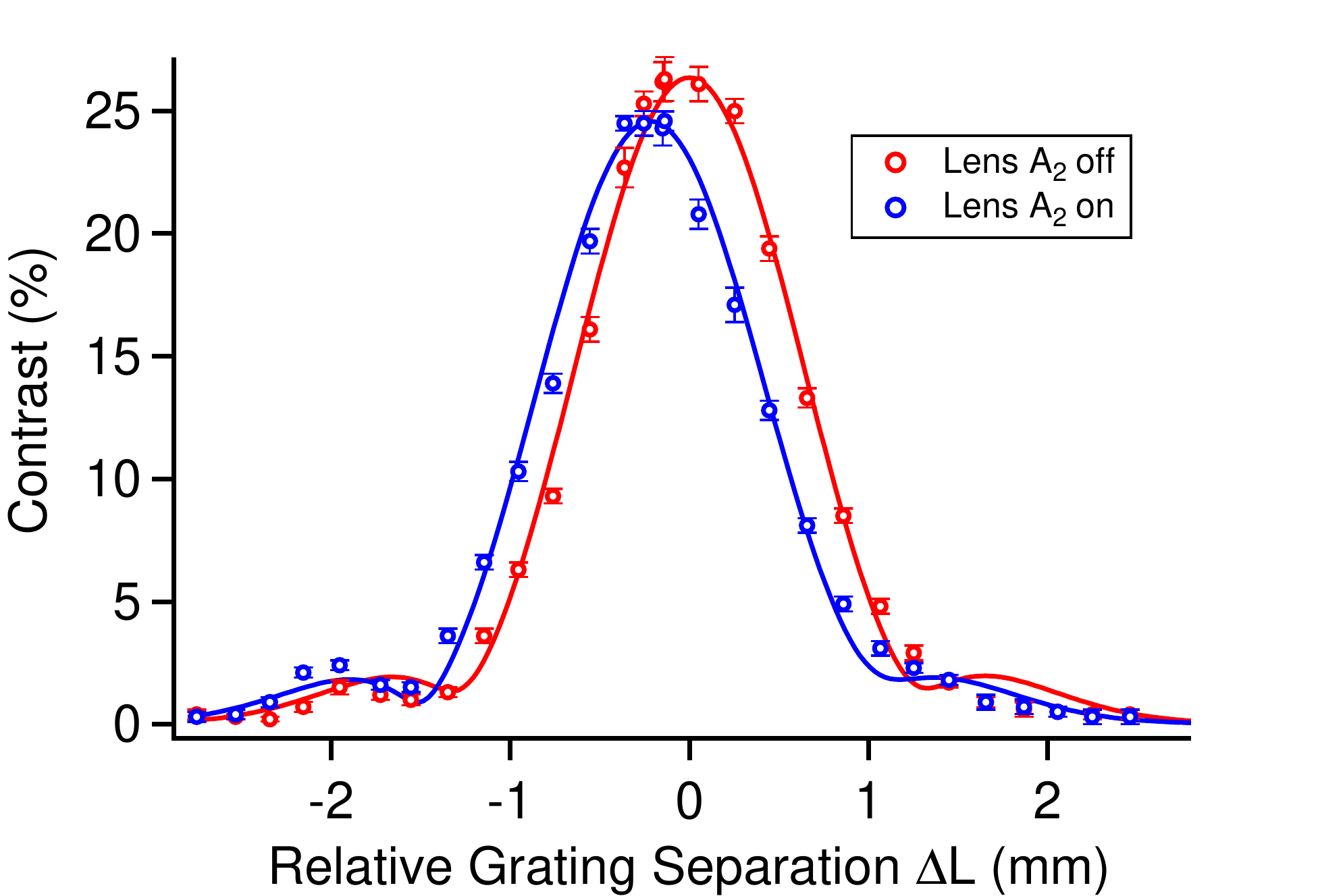}
\caption{(color online) Contrast versus $\Delta L$ measured with and without a lens. The contrast envelope shifts in position by $-205 \pm 10 ~\mu$m due to fringe magnification caused by the lens. \label{fig:CvsL}}
\end{center}
\end{figure}

In experiments we scanned the second grating in the $z$-direction to change $\Delta L$.  Figure \ref{fig:CvsL} shows contrast data as a function of $\Delta L$ with and without Lens A$_{2}$. The data exhibit peaks approximately 1.5 mm wide (FWHM in $\Delta L$).  There are also contrast revivals at $\Delta L \sim \pm 2$ mm. Importantly, the lens causes the contrast to peak at a new location shifted by \mbox{$\Delta L = -205 \pm 10$ $\mu$m}.  To interpret this shift as a measure of the focal length $f$, we will use the idea that fringe magnification due to the electrostatic lens ($M_\textrm{lens}$) can compensate for geometric magnification ($M_\textrm{geom}$).

In optics parlance, the original fringes (with no lens) are a virtual object, located a distance $|o| \approx 33$ cm to the right (i.e. downstream) of the lens.  The weak diverging lens forms a real image of the fringes at a new location, at a slightly greater distance $i$ to the right of the lens.  The image has a transverse magnification $M_T = -i/o$, and these quantities are related to the focal length ($f$) by the imaging equation ($i^{-1} + o^{-1} = f^{-1}$).  We use the convention that $o$ is negative for a virtual object, $i$ is positive for a real image, and $f$ is negative for a diverging lens, as found in several optics texts \cite{hech98,pedrotti2008introduction}.

Since the fringes are an extended object, we consider $o$ to specify particular points within the object (in a plane with a given $z$).  The image on the nanograting is actually an image of an object that was located upstream by a distance $i - |o| = |o^2/f|$, to first order in $o/f$.  That object had a smaller fringe period due to geometric magnification, so the object period was $d_o = d_g(2L_1 - |o^2/f|)/(2L_1)$.  But the image is further magnified by the lens.  Thus, before we move the nanograting from a position where $\Delta L=0$, the image has fringe period:
\begin{equation}   d_i = d_g \left( \frac{2L_1 - |o^2/f|}{2L_1} \right)  \left( \frac{o + o^2/f}{o} \right).   \end{equation}
The first factor in parenthesis accounts for the ratio of object period to grating period ($d_o/d_g$) and the second factor in parenthesis comes from the transverse magnification ($M_T$) by the lens.  Hence, the lens causes fringes at $\Delta L=0$ to be magnified, to first order in $o/f$, by
\begin{equation}   M_\textrm{lens} = 1 + \frac{o}{f} \left[1 + \frac{o}{2L_1} \right].  \end{equation}
Even a magnification of $M_\textrm{lens}=1.0005$ significantly reduces fringe contrast in our experiments because our detector is 1000 grating periods wide.

To deduce the focal length $f$ of the lens, we measured the $\Delta L$ that maximizes contrast.  We assume contrast peaks when the combined magnification $M_\textrm{lens}  M_\textrm{geom}$ equals unity.  Thus,
\begin{equation}
f = -\frac{2L_{1} o + o^2}{\Delta L}. \label{eqn:flens}
\end{equation}
Using our experimental  parameters of $L_1$ = 0.94 m, \mbox{$o = -0.33$ m}, and $\Delta L= -0.205$ mm, we measured a focal length of $f=-2.5 \pm 0.13$ km for lens A$_{2}$.  As we discuss in the next sections, this is a non-paraxial focal length, which we refer to as the Radius of Curvature $R_{c}$.

It is noteworthy that we measured such a large focal length while sampling only a 100 $\mu$m wide portion of a lens. With ordinary light optics this would be nearly impossible, because the wavelength of visible light is 100,000 times longer than the 5 pm atom waves used here.  Hence, diffraction from the beam stops would obscure any change in beam properties caused by such a weak lens.  In essence, we have monitored changes in collimation angle that are smaller than $5\times 10^{-8}$ radians, while diffraction of visible light from a 100 $\mu$m lens aperture would cause divergence on the order of $5 \times 10^{-3}$ radians.

\section{Using phase to measure $f$} \label{sec5}

\begin{figure}[h]
\begin{center}
\includegraphics[scale=0.65]{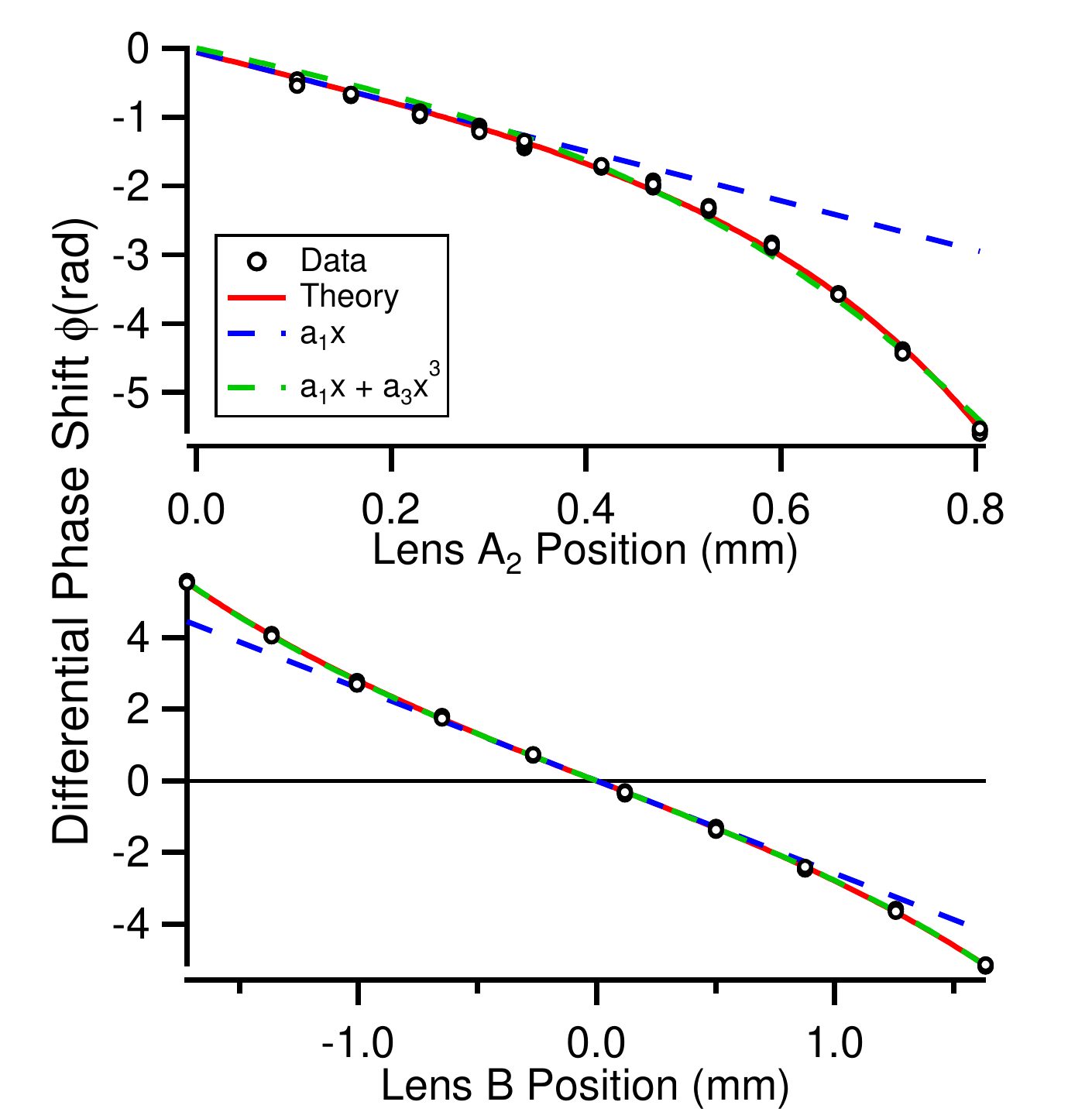}
\caption{(color online) Atom interferometer differential phase shift data. Graphs show measured values and theory for Lens A$_2$ (top) and Lens B (bottom). Differences from the linear fit shows deviation from an ideal thin lens. Error bars are smaller than the data symbols. \label{fig:ph_meas}}
\end{center}
\end{figure}

In a separate experiment, we translated the lens in the $x$-direction, perpendicular to the atom beam, and studied the differential phase shifts shown in Fig.~4.  Since both interferometer arms go through the lens, we are using our apparatus as a \textit{shearing} interferometer \cite{patorski1989self}.  If the phase induced by the lens is denoted by $\Phi(x)$ for any single path through the lens, then the observed differential phase shift $\phi$ for an interferometer with paths separated by a distance $s$ in the $x$-direction is
\begin{equation}
\phi(x) = \Phi(x+s) - \Phi(x) \label{eqn:phase}
\end{equation}

By measuring $\phi(x)$ as a function of lens position, we have mapped the derivative of $\Phi(x)$ as is typical for a shearing interferometer:
\begin{equation}
\phi(x) \approx \frac{d\Phi(x)}{dx}\cdot s. \label{eqn:dphase}
\end{equation}
Therefore a constant slope in $\phi(x)$ indicates a quadratic phase factor in $\Phi(x)$, as expected for a lens.  Furthermore, nonlinearity in $\phi(x)$ indicates spherical aberration.

We use the slope $\Delta \phi/\Delta x$ locally to find the focal length $f$ using a relationship derived for spherical waves:
\begin{equation}
f = k_g L  \left(\frac{\Delta \phi}{\Delta x}\right)^{-1} \label{eqn:fp2}
\end{equation}
where $k_g = 2\pi/d_g$ is the grating wavenumber ($d_g$ = 100 nm), and for Lens A$_{2}$, $L=-o=0.33$ m.  The observed $\Delta \phi/\Delta x$ = -3.6 rad/mm near the ground plane, so the paraxial focal length is $-5.8 \pm 0.3$  km.

To describe spherical aberration we also report how the focal length gets shorter as we use the lens farther away from its optical axis. At $x=600~\mu$m from the lens' optical axis, the slope $\Delta \phi/\Delta x$ increases to -8.8 rad/mm and the $R_{c}$ there is -2.4 km.  This is consistent with the focal length measurement for lens $A_{2}$ that we obtained in Section \ref{sec4} because $x=600 \pm 50 ~\mu$m was the position of Lens A$_{2}$ for contrast measurements in Fig.~\ref{fig:CvsL}.

We also measured $f$ for Lens B using Eq.~(\ref{eqn:fp2}). Data in Fig.~\ref{fig:ph_meas}(b) shows $\Delta \phi/\Delta x$ = -2.6 rad/mm near the optical axis. Thus the paraxial focal length for Lens B is $f= -20.0 \pm 1.0$ km. With Lens B off-axis by $x=1.15$ mm, $\Delta \phi/\Delta x$ = -3.4 rad/mm, and the $R_{c}$ (here) is $-15.3 \pm 0.8$ km.

Spherical aberration can also be quantified by fitting $\phi(x)$ with a polynomial $a_1 (x/b) + a_3 (x/b)^3$.  The best-fit parameter $a_3$ for Lens A$_{2}$ gives $W_{SA4} = a_{3}/(8\pi s) = 32 \pm 3$ waves and for Lens B gives $W_{SA4} = 130 \pm 27$ as discussed in the next section.

\section{Calculated $f$ and aberrations} \label{sec6}

In the previous two sections we presented atom optical shop testing experiments that served to measure the focal length and spherical aberrations of a lens for atoms.  Next, we calculate the focal length and spherical aberration coefficients for our atom lenses to check the measurement results.

The atom wave phase induced by our lens \cite{HHKC11,HRLC10,Rob04,Kla11,Holm13} is
\begin{equation}
\Phi(x) = \frac{4\pi\alpha V^{2}}{\hbar v} \ln^{-2}\left(\frac{a+R+b}{a+R-b}\right)\frac{b}{b^2-x^2}
\label{eqn:phi(x)}
\end{equation}
where $\alpha$ is the atomic polarizability, $V$ is the electrode voltage with respect to the ground plane, $v$ is the atom beam velocity, $x$ is the beam position relative to the optical axis and $b = a\sqrt{1+2R/a}$ is related to the geometry of the electrodes (see Table~\ref{table1}), where $R$ is the radius of the cylindrical electrode and $a$ is the electrode spacing for Lens A (or half the spacing for Lens B).

To calculate the focal length and spherical aberration coefficients of our lenses, we first find the surface of constant phase $z$, or wavefront, induced by the lens by evaluating $\Phi = - k_{dB}z$.
\begin{equation}
z(x) = - \frac{ \alpha (V/c)^{2}}{ m v^{2}}\frac{b}{\left[1-\left(x/b\right)^{2}\right]} \label{eqn:z(x)}
\end{equation}
where $c = b\left(4\pi\right)^{-1/2}\ln\left[(a+R+b)/(a+R-b)\right]$ has units of length and is comparable to the gap size in our experiments ($c=1.17$ mm for Lens A and $c=2.32$ mm for Lens B).   Written this way, $z(x)$ depends on the ratio of the potential energy $U=-\alpha E^{2}/2$ to the kinetic energy $K=mv^{2}/2$. This relation is expected since the index of refraction for atom waves due to the electric field is $n=(1-U/K)^{1/2}$ \cite{ASM94}; hence $n-1$ depends to first order linearly on the small parameter ($U/K$). Therefore, we introduce a dimensionless parameter $g = \alpha (V/c)^{2}/(mv^2)$ for what follows.  In our experiments $g \approx 10^{-7}$, which explains why the focal lengths are so long for our electrostatic lenses for atom beams.

The radius of curvature, $R_{c}$ of the iso-phase surface $z$ can be found from a local circle fit to Eq.~(\ref{eqn:z(x)}) (the so-called osculating circle):

\begin{equation}
R_{c} =  - \frac{b}{2g}  \frac{ \left\{ [1-(x/b)^2]^4 - 4g^2(x/b)^2 \right\}^{3/2} }{ [1+3(x/b)^2]  [1-(x/b)^2]^3}. \label{eqn:ROC} \end{equation}
Equation (\ref{eqn:ROC}) also shows how the radius of curvature $R_c$ depends on the distance from the optical axis $x$ (spherical aberration) and the de Broglie wavelength $\lambda_{dB}$ since $g$ depends on $v^{-2}$ (chromatic aberration).  At $x=0$, i.e. on the optical axis, the focal length is equal to $R_{c}$, hence
\begin{equation}
f = -\frac{b}{2g}
\label{eq:fl}
\end{equation}
For a 2000 m/s K atom beam, we calculate $f = - 6.1$ km for Lens A$_{2}$ and $f = - 21.7$ km for Lens B. These calculations are in agreement with the focal length measurements. For Lens A$_{2}$ at $x=600$ $\mu$m,  $R_{c} = - 2.3$ km. This calculation is consistent with the measurements of $R_{c}$ in Sec. \ref{sec5}.

Next, we study $z(x)/\lambda_{db} = W_{A}$, known as the aberrated wavefront, which we have expressed in units of waves. We proceed by expanding $W_{A}$  in a Taylor series as is done in light optics \cite{greivenkamp2004field,sasian2012introduction}.

\begin{equation}
W_{A} = - \frac{bg}{\lambda_{dB}}\left[1+\left(\frac{x}{b}\right )^{2} + \left(\frac{x}{b}\right )^{4} + \cdots \right ]
\label{eqn:taylor}
\end{equation}
The zeroth order term in $x/b$, the first term in brackets, is position-independent and corresponds to an optical flat with no optical power. The second order term is the focusing term. It corresponds to an ideal lens with a focal length $f = -b/(2g)$, same as the paraxial focal length in Equation (\ref{eq:fl}).

To determine the spherical aberration, we first calculate the wavefront error $W = W_{A}-W_{R}$, where $W_{R}$ is an unaberrated reference spherical wavefront with a radius of curvature equal to the paraxial focal length. For our lenses, the focal length is so much greater than the lens aperture that the difference between our spherical reference wavefront and a parabolic wavefront is negligible. Therefore, for our lenses,  $W \approx W_{A}$ for higher order terms in $x/b$.

Terms that are 4th order and higher in $x/b$ describe the different order spherical aberrations.  For our electrostatic lenses, the wavefront spherical aberration coefficients are the same for all orders of spherical aberration. We therefore refer to the coefficients as $W_{SA}$. For our lenses,
\begin{equation}
W_{SA} = - \frac{bg}{\lambda_{dB}}
\end{equation}
For Lens A$_{2}$ and a 2000 m/s K atom beam, we compute $W_{SA} = -35$ waves. Similarly, for Lens B we compute $W_{SA}= -130$ waves. This indicates that 1.12 mm from the optical axis for Lens B, there is only a 1/4 de Broglie wave (1 pm) deviation between $W_{A}$ and $W_{R}$.  These calculations are in agreement with the measurements of the fourth order spherical aberration coefficient $W_{SA4}$ for Lens A$_{2}$ and Lens B in Sec.\ref{sec5}.

There is also axial chromatic aberration and spherochromatism in our electrostatic lenses \cite{greivenkamp2004field} because the focal length depends on $\lambda_{dB}^{-2}$ and the spherical aberration coefficients are proportional to $\lambda_{dB}^{-3}$.

\section{Application for Velocity measurement} \label{sec7}

\begin{figure}[h]
\begin{center}
\includegraphics[width=9cm]{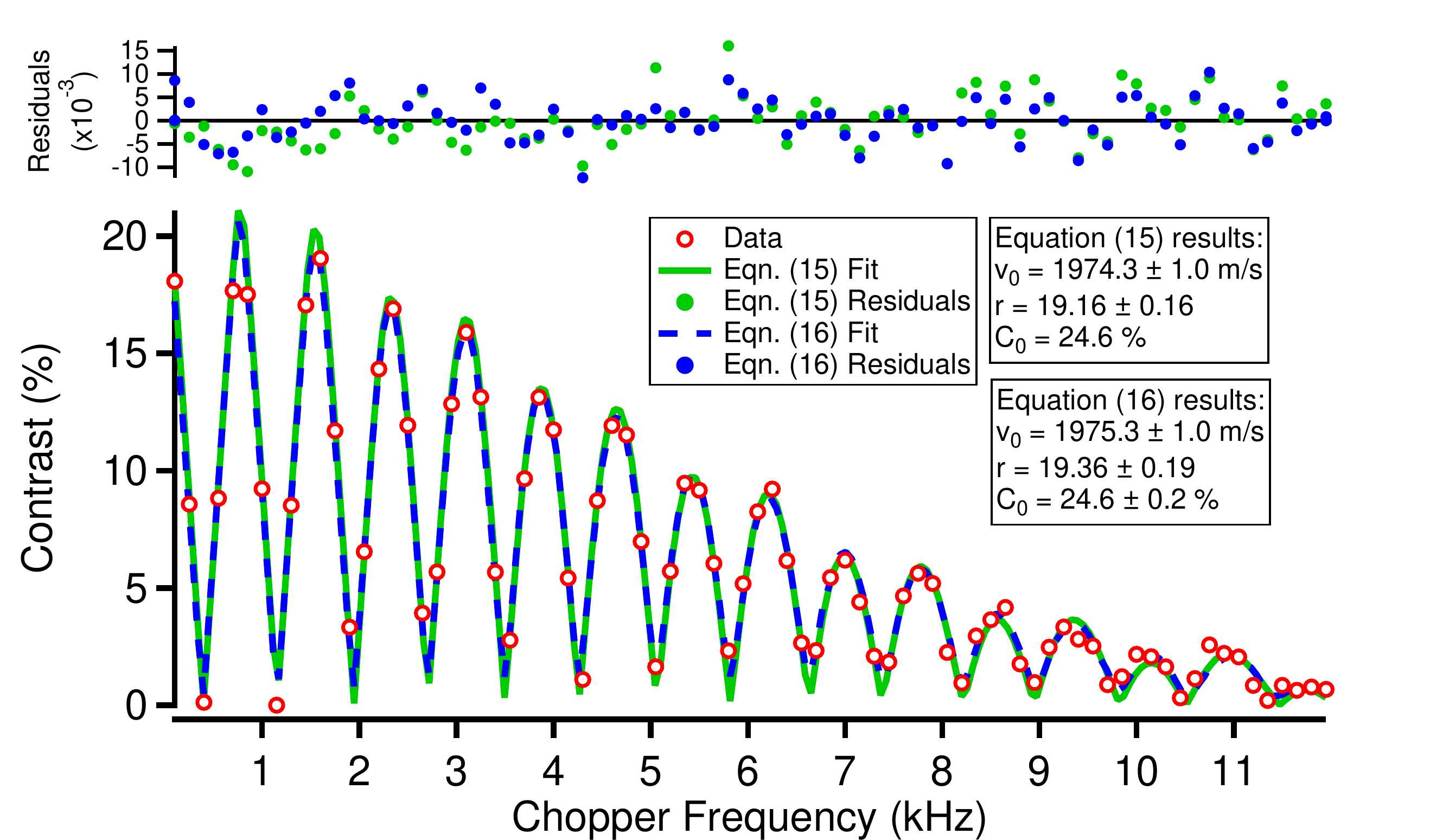}
\caption{(color online) Contrast data versus chopping frequency $C(\nu)$ analyzed with Eq.~(\ref{newV}) and Eq.~(\ref{oldV}). The best fit parameters for the most probable velocity and velocity ratio ($v_0$ and $r$) are discussed in the text. \label{chopdata}}
\end{center}
\end{figure}

Defocusing by phase chopper electrodes (Lens A$_1$ and Lens A$_2$) affects the accuracy of our atom beam velocity measurements. Although we have made mean velocity measurements using phase choppers within 0.05$\%$ statistical precision \cite{Holm13}, a systematic error of approximately 0.1\% can be attributed to defocusing.  Furthermore, this error changes with $\Delta L$. We learned to recognize and fix this problem by viewing electric field gradients as lenses for atom waves.

Holmgren \textit{et. al.} \cite{HHKC11} described how phase choppers are used to measure the velocity of atom beams.  In brief, phase choppers are analogous to mechanical choppers.  But fringe contrast, rather than beam flux, is modulated as a function of chopping frequency, $\nu$.  We analyze $C(\nu)$ data to measure the mean velocity and the velocity spread of our atom beam, as shown in Figure \ref{chopdata}.  The mean velocity determines the frequencies at which the contrast revivals occur, and the velocity spread affects how the contrast revivals decay.

In equation (1) of reference \cite{HHKC11} we modeled the contrast and phase of the interference fringes by
\begin{align}
C(\nu)e^{i\phi(\nu)} &= C_{0}e^{i\phi_{0}}\nu \int_{t=0}^{1/ \nu} \int_{v=0}^{\infty}P(v)\nonumber \\
 & \times e^{i[\phi_{1}(v,t)+\phi_{2}(v,t+\ell/v)]}dvdt \label{oldV}
\end{align}
where $\phi_{1}$ and $\phi_{2}$ were the $v$-dependent (but $x$-independent) phase shifts due to choppers $A_{1}$ and $A_{2}$, and $P(v)$ is the probability distribution for velocity.  We will use a gaussian $P(v) = (2\pi)^{-1/2}\exp{[-(v-v_0)^2/(2 \sigma_v^2)]}$ and keep both $v_0$ and the ratio $r=v_0/\sigma_v$ as free parameters when comparing Eq.~(\ref{oldV}) to $C(\nu)$ data.

Our analysis of lenses inside an atom interferometer helped us develop an improved model that includes the thickness of the atom beam and its angular spread by explicitly averaging over all detected trajectories.  We do this by integrating over all transverse positions in the two collimating slits.  We also include the transverse coherence length of the atom beam, the symmetric pairs of interferometers sketched in Figure 1, and the finite size of the atom beam detector. We thus replace Eq.~(\ref{oldV}) with
\begin{align}
C(\nu)e^{i\phi(\nu)}&=  C_0 e^{i\phi_{0}}  \frac{1}{2} \sum_{j=-1,1}
\int_{-x_{1}/2}^{+x_{1}/2}     \int_{-x_{2}/2}^{+x{_2}/2}  \nonumber \\
& \times \nu \int_{t=0}^{1/\nu} \int_{v=0}^{\infty}P(v)  D_j(x_{1},x_{2},v) \nonumber \\
& \times C_{\textrm{env}}(\Delta L,t) e^{i\phi_{\textrm{sep},j}(x_1, x_2, v, \Delta L)}  \nonumber \\
& \times e^{i[\phi_{1j}(x_{1}, x_2,v,t)+ \phi_{2j}(x_1,x_2,v,t+\ell/v)]} \nonumber \\
& \times e^{i\phi_{\textrm{sag}}(v)}dv dt dx_{1} dx_{2} \label{newV}
\end{align}
where $C_0$ is the reference contrast observed when $\Delta L = 0$ and the electrodes are grounded.  The sum over $j$ is for interferometers formed by different pairs of diffraction orders (see Figure 1). $D_j(x_{1},x_{2},v)$ describes the probability that atoms hit the detector after passing through positions $x_1$ and $x_2$ in the two collimating slits and diffract in the directions given by $j$ and $v$.  $C_{\textrm{env}}(\Delta L,t)$ is a contrast envelope due to transverse coherence length, which we modeled with GSM beams and has been discussed in \cite{CBV99,miffre2005lithium}. The Sagnac phase is $\phi_{\textrm{sag}}$ \cite{HRLC10}. We model $\phi_{\textrm{sep}}$, the separation phase \cite{DGHK08}, as
\begin{equation}
\phi_{\textrm{sep}}(\Delta L)= k_g \left(\theta_{\textrm{inc}}(x_1,x_2) + j \frac{\lambda_{dB}}{2d_{g}}\right)\Delta L \label{eqn:psep2}
\end{equation}
where $\theta_{\textrm{inc}}$, the angle of incidence on the first grating, plus (or minus) half the diffraction angle ($\lambda_{dB}/2d_{g}$) represents the angle of the fringe maxima in the $x$-$z$ plane (see Fig.~2).  This angle depends on $x_1, x_2, v$, and $j$, and is summarized by the idea of geometric magnification.  Since neither $C_{\textrm{env}}$ nor $\phi_{\textrm{sep}}$ were included in Eq.~(\ref{oldV}), it did not depend on $\Delta L$.

The terms $\phi_{ij}$ in Eq.~(\ref{newV}) with $i\in(1,2)$ represent the phase imparted by chopper $i$ to atoms that fly by with wave function components at transverse positions $x_{ij}$ and $x_{ij}+s$ based on Eq.~({\ref{eqn:phi(x)}).  These positions depend on $x_1$, $x_2$, $v$, and $j$, therefore we integrate and sum over $\phi_{ij}$ as well.  These position-dependent phases, $\phi_{ij}$, take into account that the phase choppers are lenses for atoms.

Using Eq.~(\ref{oldV}) or  Eq.~(\ref{newV}) to make a least-squares fit to data in Fig.~\ref{chopdata} results in slightly different parameter values for $v_0$ and $r$.  The values of $v_0$ using the two theoretical models are different by about 0.2\% for slow atom beams ($v_{0} < 1000$ m/s).  To investigate which analysis is more accurate, we were motivated to test other predictions of focusing behavior, such as $C(\Delta L)$ data shown in Fig.~3 and $\phi(x)$ data shown in Fig.~4 as further evidence that Eq.~(\ref{newV}) is indeed more accurate than Eq.~(\ref{oldV}).

Next, we examine model-dependent differences in the best fit parameters $v_0$ and $r$ as a function of $\Delta L$. We used Eq.~(\ref{newV}) to simulate $C(\nu)$ spectra with particular $v_0$ and $r$ given as input parameters.  Then, we used the simpler model described by Eq.~(\ref{oldV}) to find the best fit parameters $\tilde{v_0}$ and $\tilde{r}$ that minimize the sum of the squared errors when analyzing the simulated data.  The differences $\tilde{v_0}- v_0$ and $\tilde{r}-r$ depend on $\Delta L$ as shown in Fig.~\ref{vvsl}.  This is significant because $\Delta L$ was not even considered as a parameter in earlier work on phase choppers \cite{Holm13,HHKC11,Kla11}, but Fig.~\ref{vvsl} now shows that it is important.

\begin{figure}[t]
\begin{center}
\includegraphics[scale=0.5]{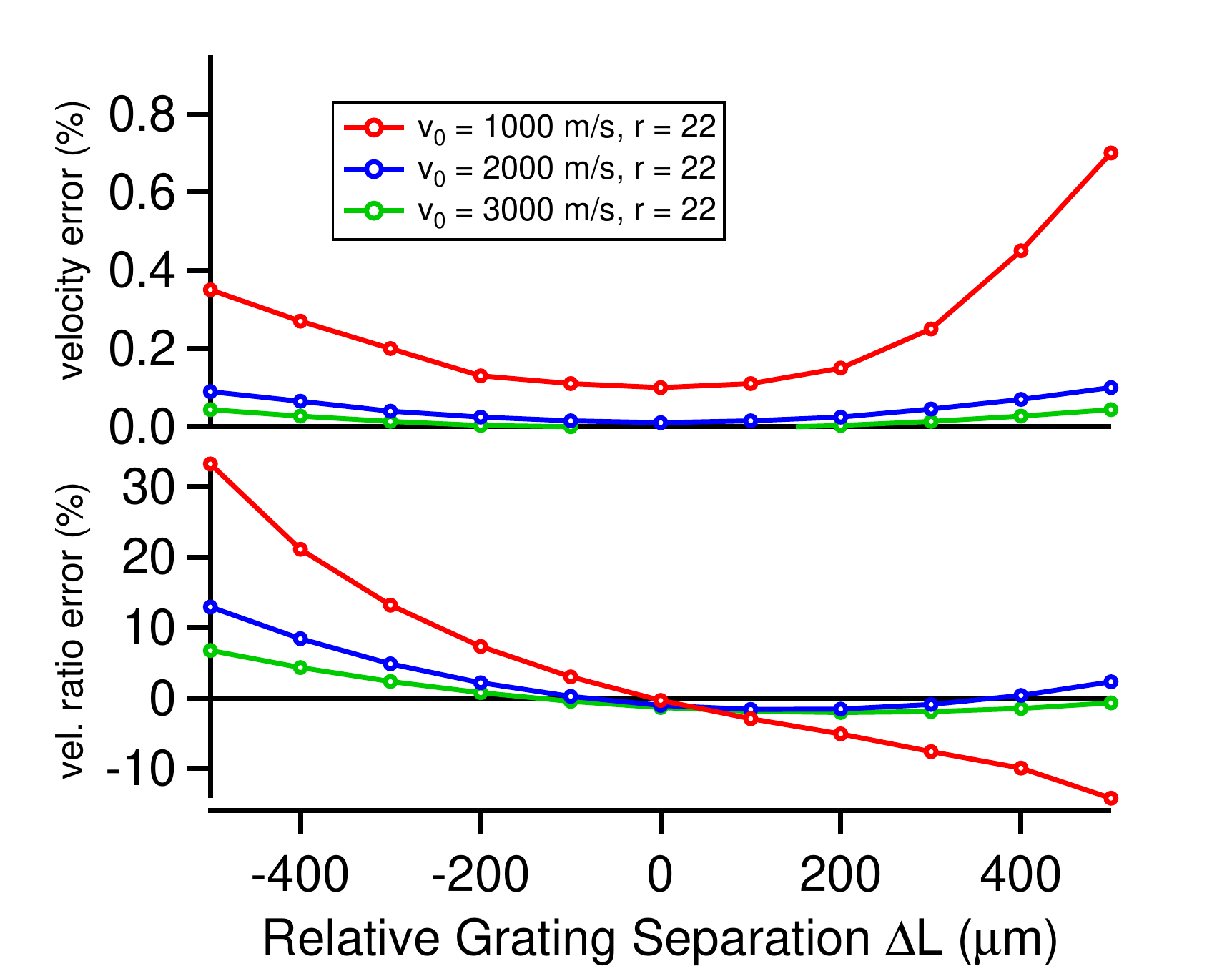}
\caption{(color online) Error in best fit velocity as a function of $\Delta L$ using Eq.~(\ref{oldV}) to fit simulated data generated with the more complete model described by Eq.~(\ref{newV}).  The velocity error is $(\tilde{v_0}-v_0)/v_0$ and the error in velocity ratio is $(\tilde{r}-r)/r$ as described in the text.
\label{vvsl}}
\label{fig:fig5}
\end{center}
\end{figure}

Most trends in Fig.~\ref{vvsl} can be explained by fringe magnification. For example, reduction in contrast due to focusing can be misinterpreted as a larger spread in velocity.  In particular, if $\Delta L > 0$ then focusing reduces the contrast,  so Eq.~(\ref{oldV}) makes a best fit $\tilde{r}$ too low. Conversely, when $\Delta L < 0$ choppers can increase contrast, as we saw in Fig.~3, and as a consequence the best fit $\tilde{r}$ is too large.  Hence, the combined influence of $\phi_{ij}$ and $\phi_{sep}$ (electrostatic and geometric magnification) causes errors in $\tilde{r}$ that can be anti-correlated with $\Delta L$, as seen in Fig.~\ref{vvsl}.  Velocity-dependent contrast suppression and velocity-dependent detection included in Eq.~(\ref{newV}) but not Eq.~(\ref{oldV}) also influence errors in $\tilde{r}$.

Trends in $\tilde{v_0}$ are similarly subtle but can also be explained, for the most part, in terms of magnification.  One reason for an error in $\tilde{v_0}$ is that slow atoms, with a larger diffraction angle, preferentially miss the detector.  This is described by $D_j(x_{1},x_{2},v)$.  Just as importantly, when the slower atoms in the ensemble contribute to the signal with low contrast as a result of a mechanism not adequately described by Eq.~(\ref{oldV}), then there will be a bias towards faster $\tilde{v_0}$.  Preferential contrast loss for slow atoms is caused, for example, by geometric magnification ($\phi_{sep}$) simply because of the larger diffraction angles.  Also, electrostatic magnification ($\phi_{ij}$) generally produces velocity-dependent contrast-loss due to chromatic aberration.  When $\Delta L$ is slightly negative, the signal can be biased towards slower atoms because electrostatic magnification then compensates for geometric magnification preferentially for the slow atoms.

We have documented our effort to validate the model in Eq.~(\ref{newV}) with auxiliary tests, e.g. in Fig.~3 and Fig.~4, so we are convinced that Eq.~(\ref{newV}) is an improvement over Eq.~(\ref{oldV}).  Furthermore, in case there are still additional position- and velocity-dependent phases that we have not yet discovered that make an impact on our analysis, we can offer a strategy to minimize errors in measurements of $v_0$ and $r$ by operating at a $\Delta L$ where errors are minimum.  We note there are minima in the absolute value of $\tilde{v_0}$ error and $\tilde{r}$ error in Fig.~\ref{vvsl} when $\Delta L  =  -100 ~\mu$m.  The reason for a minimum error is because at this $\Delta L$ the contrast change due to the phase choppers is very small for atoms with $v_{0}$.  Recall, this is where the two contrast curves intersect in Fig.~3.  Therefore we recommend operating future phase chopper experiments with $\Delta L$ chosen so the observed contrast is not affected by the phase choppers.  Then, even a simpler model that is missing some contrast-reducing mechanisms [e.g. Eq.~(\ref{oldV})] still produces a smaller error in $\tilde{v_0}$ and $\tilde{r}$ than it would if $\Delta L=0$.

\section{Why focusing is unavoidable} \label{sec8}

Next we address the question: Can electrodes be fabricated with shapes that produce zero focusing, but still cause deflection?  If a prism from atom waves were possible, then more ideal phase chopper could be constructed and some of the analysis described by Eq.~(\ref{newV}) could be avoided.  However, in this section we conclude that focusing is unavoidable if we use electrodes that are invariant under translation in the $y$-direction, and static (or low frequency) electric fields.

We pose this question mathematically by asking if there exists a vector field $\mathbf{E}(x,z)$ that produces zero focusing power:
\begin{equation}   \frac{d}{dx} \phi = \left(\frac{\alpha s}{2\hbar v}\right) \frac{d^2}{dx^2} \int \mathbf{E}^2 dz = 0, \label{eq:d2p} \end{equation}
but also makes non-zero deflection:
\begin{equation}  \phi =  \left(\frac{\alpha s}{2\hbar v} \right) \frac{d}{dx} \int \mathbf{E}^2 dz \neq 0.  \label{eq:d1p} \end{equation}
Here $\alpha$ is the atomic polarizability, $s$ is the separation between paths in the atom interferometer, $\hbar$ is Planck's constant, and $v$ is the atomic velocity.  We can rewrite Eq.~(\ref{eq:d2p}) to be
\begin{equation}
\frac{d\phi}{dx} = \left(\frac{\alpha s}{\hbar v} \right) \int \left[\left(\frac{d}{dx}\mathbf{E}\right)^2  + \mathbf{E} \frac{d^2}{dx^2}\mathbf{E} \right] dz. \label{dpdx}
\end{equation}
Then, since $\nabla^2 \mathbf{E} = 0$, for $y$-invariant fields we know \mbox{$\frac{d^2}{dx^2}\mathbf{E}  = - \frac{d^2}{dz^2}\mathbf{E}$}. Thus we can replace the second term in Eq.~(\ref{dpdx}) with $ -\mathbf{E} \frac{d^2}{dz^2}\mathbf{E} $ and integrate by parts to show
\begin{equation}
\frac{ d\phi}{dx} = \left( \frac{\alpha s}{\hbar v} \right) \int \left[\left(\frac{d}{dx}\mathbf{E}\right)^2  + \left(\frac{d}{dz}\mathbf{E}
\right)^2 \right] \label{eq:pos} dz.
\end{equation}

This expression for the focusing power is positive definite.  Thus, if there is any gradient in the electric field, as there must be to cause a deflection, then Eq.~(\ref{eq:pos}) shows that the focusing power is nonzero.  This proves that Eq.~(\ref{eq:d2p}) and Eq.~(\ref{eq:d1p}) cannot both be satisfied.

\section{Discussion} \label{sec9}

In summary, we conducted atom optical shop testing experiments using an atom interferometer to measure the focal lengths of lenses for atoms.  Measurements of shearing interferometer phase shifts $\phi(x)$ induced by a lens enabled us to report $f$ with less than 5\% uncertainty even though the values of $f$ that we studied were quite large (ranging from -2.5 km to -21.7 km).  We also used $\phi(x)$ to measure spherical aberration.   Measurements of $C(\Delta L)$ showed how electrostatic magnification of atom interference fringes can compensate for geometric magnification.  This observation explained systematic shifts in measurements of atomic beam velocity $v_0$ and velocity spread $r$ made with phase choppers.  Then, an improved model of $C(\nu)$ was developed to reduce these systematic errors by properly representing geometric and electrostatic magnification.  Systematic corrections were shown to depend on $\Delta L$, a parameter that had not previously been considered in the analysis of phase choppers.

The goal of this improved model of phase chopper $C(\nu)$ spectra is to support future measurements of atomic polarizability.  For this application we recommend monitoring $\Delta L$ using contrast measurements and incorporating uncertainty in $\Delta L$ into the error budget for resulting measurements of atomic velocities, velocity ratios, and ultimately polarizabilies. Additional modifications to Eq.~(\ref{newV}) may be explored, such as diffraction phases induced by the gratings themselves \cite{Per06}, but we suspect that diffraction phases in particular can be incorporated simply by using the contrast peak to redefine where $\Delta L=0$.  Thus, we recommend measuring $\Delta L$ with respect to the contrast peak and then selecting a $\Delta L$ that makes the phase choppers cause minimal changes in contrast.

To give a broader perspective, electron and optical interferometers routinely use lenses to intensify, magnify, and focus fringes. Neutron optics experiments \cite{rauch2000neutron} have also begun to include magnetic hexapole  \cite{oku2000neutron} or solid state lenses \cite{kumakhov1992neutron,beguiristain2002simple} as tools to manage neutron beams.  Therefore, given the many examples of atom lenses discussed in the literature  \cite{kaenders1996refractive,jardine2001hexapole, CLB06,ketterle1992trapping, G55,noh2000imaging,KAG05, CSSTM91,rdg00, BFAP78,SPBM92,mcclelland1995atom, DGR99, KRS08, bam99,TBTCPB92,mcclelland1993laser},  we expect several new opportunities may result from using lenses for atoms in conjunction with an atom interferometer.  In this paper we used this nexus to develop atom optical shop testing, in analogy to ordinary optical shop testing \cite{Malacara07, greivenkamp2004field, sasian2012introduction, goodwin2006field, hech98, pedrotti2008introduction}.  We also explored systematic corrections to precision measurements that arise when de Broglie wave curvature is manipulated inside an atom interferometer.

The analogy of a lens in an interferometer may be extended to ultracold atom interferometers operated in traps. However, in traps the parameter $g=U/K$ is often larger than 1 and the interaction time with the trap potential can be many times the inverse trap frequency.  In comparison, for experiments presented here $g\approx 10^{-7}$, the fly-by time was a few $\mu$s, and the resulting the displacement of atoms within the lens was negligible compared to the atom beam width or even its transverse coherence length.  Thus, the electric field gradients we studied are like weak, diverging, cylindrical, thin lenses for atoms, whereas traps are more similar to waveguides for atoms~\cite{Meystre2001atom,folman2002microscopic,CSJD09}.

Finally, we speculate on more uses for lenses in atom interferometers.  Schemes may be developed to magnify atom interference fringes to make them easier to image.  We also propose lenses can compensate for gratings that have the wrong period.  For example, if the nanogratings in a hybrid KD-TLI interferometer \cite{gerlich2007kapitza} are imperfectly matched with the laser grating period, then atom lenses may improve the fringe contrast.  Spherical aberration can compensate for grating chirp.  Strong positive lenses can focus beams inside atom interferometers to increase flux and reduce beam widths.  Atom optical shop testing with an atom interferometer could also validate aberration mitigation techniques such as those proposed in \cite{ICAP91} and \cite{EMP13}.

\section{Acknowledgments}
This work is supported by NSF Grant No.~1306308 and a NIST PMG.  Authors R.T. and M.G. thank the NSF GRFP Grant No. DGE-1143953 for support. Author W.F.H. thanks the Arizona TRIF for additional support.

\appendix

\bibliography{lensbib}
\end{document}